\documentclass[preprint, prb, aps, superscriptaddress, endfloats*]{revtex4-1}

\usepackage{times}
\usepackage{mathptmx}
\usepackage{amsmath}
\usepackage{upgreek}
\usepackage{graphicx}
\usepackage{array}
\usepackage{lineno}
\usepackage{natmove}
\usepackage{hyperref}
\usepackage{xcolor}
\definecolor{linkcolor}{RGB}{0,0,150}
\hypersetup{%
  colorlinks,
  urlcolor    = linkcolor,
  linkcolor   = linkcolor,
  citecolor   = linkcolor
}

\newcommand{\ie}[0]{i.e.}
\newcommand{\eg}[0]{e.g.}
\newcommand{\etal}[0]{et al.}

\begin{document}



\title{Effects of bulk and interfacial anharmonicity on thermal conductance at solid/solid interfaces}%

\author{Nam Q.~Le}%
\email{nql6u@virginia.edu}
\altaffiliation{Present Address: US Naval Research Laboratory, Washington, DC 20375}
\affiliation{Department of Mechanical and Aerospace Engineering, University of Virginia, Charlottesville, Virginia 22904}

\author{Carlos A.~Polanco}%
\affiliation{Department of Electrical and Computer Engineering, University of Virginia, Charlottesville, Virginia 22904}

\author{Rouzbeh Rastgarkafshgarkolaei}%
\affiliation{Department of Mechanical and Aerospace Engineering, University of Virginia, Charlottesville, Virginia 22904}

\author{Jingjie Zhang}%
\affiliation{Department of Electrical and Computer Engineering, University of Virginia, Charlottesville, Virginia 22904}

\author{Avik W.~Ghosh}%
\affiliation{Department of Electrical and Computer Engineering, University of Virginia, Charlottesville, Virginia 22904}
\affiliation{Department of Physics, University of Virginia, Charlottesville, Virginia 22904}

\author{Pamela M.~Norris}%
\email{pamela@virginia.edu}
\affiliation{Department of Mechanical and Aerospace Engineering, University of Virginia, Charlottesville, Virginia 22904}

\date{\today}


\begin{abstract}
  We present the results of classical molecular dynamics simulations to assess the relative contributions to interfacial thermal conductance from inelastic phonon processes at the interface and in the adjacent bulk materials. The simulated system is the prototypical interface between argon and ``heavy argon'' crystals, which enables comparison with many past computational studies. We run simulations interchanging the Lennard-Jones potential with its harmonic approximation to test the effect of anharmonicity on conductance. The results confirm that the presence of anharmonicity is correlated with increasing thermal conductance with temperature, which supports conclusions from prior experimental and theoretical work. However, in the model Ar/heavy-Ar system, anharmonic effects at the interface itself contribute a surprisingly small part of the total thermal conductance. The larger fraction of the thermal conductance at high temperatures arises from anharmonic effects away from the interface. These observations are supported by comparisons of the spectral energy density, which suggest that bulk anharmonic processes increase interfacial conductance by thermalizing energy from modes with low transmission to modes with high transmission.
\end{abstract}

\maketitle

\section{Introduction}
\label{sec:introduction}

The contribution of inelastic phonon processes to the thermal conductance at solid/solid interfaces is a topic of enduring interest. At interfaces between metal films and dielectric substrates whose phonon spectra are extremely mismatched---\eg,~Pb/diamond---experimentally measured values can far exceed the phonon radiation limit~\cite{Stoner1992,Stoner1993,Lyeo2006,Hohensee2015}, which represents the upper limit of conductance when accounting only for elastic (\ie,~frequency-preserving) phonon transmission. The measured values also increase monotonically with temperature, in common with calculations of conductance from molecular dynamics (MD) simulations which naturally include anharmonic effects~\cite{Stevens2007,Landry2009a,Saaskilahti2014a}. These observations strongly suggest that inelastic scattering (\ie,~energy transfer among modes of different frequency) contribute a large fraction of conductance at high temperature. Since inelastic processes arise from anharmonicity of interatomic forces, the contribution is also expected to grow as temperature (and hence atomic displacement) increases, making it relevant to thermal engineering in applications with high operating temperatures such as high-power electronics~\cite{Sarua2007,Kuzmik2007,Mishra2008}.

The seminal models for predicting conductance, the acoustic mismatch model~\cite{Khalatnikov1952,Little1959} and diffuse mismatch model (DMM)~\cite{Swartz1987,Swartz1989}, only account for elastic transmission processes. Only elastic processes are expected in a system with harmonic interatomic forces or, alternatively, in an anharmonic system under small displacements. Under this assumption, the DMM provides a first approximation for estimating the conductance. Based on comparison with experimental data, the DMM appears to generally overestimate the conductance between vibrationally well-matched materials and underestimate the conductance between mismatched materials~\cite{Stevens2005}. The degree of matching is often summarized in the ratio of Debye temperatures, $\theta_\text{D}$, and the transition between these two regimes is observed empirically when $\theta_\text{D}$ of the substrate is $\sim$3--4 times that of the film~\cite{Stevens2005}. 
For example, lead and diamond have an extraordinarily high mismatch in vibrational spectra: the highest-frequency phonons in Pb are $\sim$2.2~THz, while those in diamond are $\sim$39.2~THz~\cite{Brockhouse1962,Warren1967}. The expected conductance due only to \emph{elastic} phonon transmission is correspondingly low, on the order of 2~MW~m$^{-2}$~K$^{-1}$. However, this underestimates experimental measurements by a full order of magnitude, with reported values ranging roughly 20--60~MW~m$^{-2}$~K$^{-1}$~\cite{Stoner1993,Lyeo2006,Hohensee2015}.

Several modifications to the DMM have been proposed to account for inelastic transmission processes in predictions of conductance~\cite{Kosevich1995,Hopkins2007,Hopkins2009d,Hopkins2011a,Duda2011}. For example, Hopkins and coworkers proposed two modifications to the DMM\@: the higher harmonic inelastic model (HHIM)~\cite{Hopkins2009d} and the anharmonic inelastic model (AIM)~\cite{Hopkins2011a} which provide expressions for the transmissivities corresponding to $n$-phonon processes: $\omega_1 + \omega_2 + \cdots + \omega_{n-1} \leftrightarrow \omega_n$, where $\omega$ denotes phonon frequency. By comparison an elastic (2-phonon) process would be denoted $\omega \to \omega$. The HHIM only allows processes that combine phonons of equal frequency ($\omega_1 = \cdots = \omega_{n-1}$), while the AIM allows the combination of phonons of arbitrary frequency. Duda and coworkers also proposed a modification to the DMM that incorporates bulk-like scattering near the interface rather than at the interface itself, which they used to predict an increasing conductance with temperature in the classical limit~\cite{Duda2011}. Despite making different assumptions about the details of inelastic processes, these models improve agreement with conductance measurements to similar degrees, making it difficult to determine their relative validity.

Several recent works have elucidated the details of inelastic processes and their contributions to thermal conductance. The theoretical and computational work by S\"{a}\"{a}skilahti~\etal~\cite{Saaskilahti2014a} showed that frequency-doubling and -halving processes dominate the inelastic contribution to conductance in MD simulations, lending support to the assumptions of the HHIM\@. However, Hohensee~\etal~\cite{Hohensee2015} observed experimentally that the conductance of metal/diamond interfaces depends only weakly on pressure, from which they inferred that inelastic processes involving two metal phonons of equal frequency cannot be the dominant contribution to the conductance. Both works observed that their conclusions may be reconciled by careful consideration of the inelastic processes in the bulk-like regions near the interface. This precise question was investigated in Refs.~\onlinecite{Wu2014} and~\onlinecite{Murakami2014}. Using MD, Wu and Luo~\cite{Wu2014} simulated the conductance between one crystal with a monatomic basis and another crystal with a diatomic basis. They observed that increasing an anharmonic force constant in the diatomic lattice increased the total conductance dramatically due to increased coupling between acoustic and optical modes. By contrast, increasing an anharmonic force constant of the interfacial interaction had no effect on the conductance. This is broadly consistent with our results, but differs with our observation that the interfacial contribution is significant (though smaller than the bulk contribution). Furthermore, the present results expand on how the relative contributions change with temperature. A related difference is that size effects were observed, which did not affect their qualitative conclusions but precluded the quantitative comparison of the contributions from elastic, bulk inelastic, and interfacial inelastic processes. Nevertheless, we observe the same general mechanism that Wuo and Luo identified: inelastic processes contribute to conductance via the bulk thermalization of modes with low transmissivity. In a different work, Murakami~\etal~\cite{Murakami2014} made related conclusions from MD simulations of PbTe/PbS and Si/Ge interfaces, in which they demonstrated the importance of inelastic processes in a broad transition region (TR) rather than only at the plane of the interface. Inelastic processes in the TR downconvert energy from high-frequency to low-frequency modes, which then transmit elastically, in agreement with our observations. However, their analysis did not provide a direct calculation of the separate elastic and inelastic contributions to conductance, nor the temperature dependence of the contributions, which will be essential for testing models that correctly incorporate inelastic processes.

Therefore, the goal of the present work is to decompose the thermal conductance at a model interface into explicit contributions from the harmonic dynamics, the anharmonic effects at the interface, and the anharmonic effects in the bulk materials. Our model system is a planar interface between Ar and ``heavy Ar,'' which has been the prototypical model system for studying these phenomena. In Sec.~\ref{sec:nemd} we present calculations of the conductance in the model system with different configurations of harmonic and anharmonic forces between atoms. The results confirm that conductance rises with temperature only in the systems with anharmonic forces, which presumably enable inelastic phonon processes. However, at high temperatures, the anharmonicity at the interface itself appears to contribute less than half of the total conductance in our model system; the anharmonicity in the bulk materials is responsible for the rest. 
These observations are corroborated in Sec.~\ref{sec:wavelet}, in which use the wavelet transform to calculate the spectral energy densities throughout the interfacial systems. Those spectra show that energy reflected from the interface is in strong non-equilibrium, and anharmonicity enables its thermalization, suggesting a mechanism to explain the increase in interfacial conductance. We summarize the findings in Sec.~\ref{sec:conclusions} and comment on their relation to other research on this topic.

\section{Effect of local anharmonicity on interfacial thermal conductance}
\label{sec:nemd}

In this section, we present calculations of interfacial thermal conductance using non-equilibrium molecular dynamics (NEMD). Further simulation details are given in Appendix~\ref{sec:appendix_ar}. As a prototypical anharmonic potential, we use the Lennard-Jones (LJ) potential $U_\text{LJ}(r_{ij}) = 4 \epsilon \left[ (\sigma/r_{ij})^{12} - (\sigma/r_{ij})^6 \right]$ where $r_{ij}$ is the distance between atoms $i$ and $j$, $\epsilon$ is the energy scale, and $\sigma$ is the length scale. The LJ potential is strongly anharmonic, which induces inelastic phonon processes. In order to suppress inelastic phonon processes in certain regions, we replace the LJ potential with its second-order Taylor expansion about the equilibrium separation $r_\text{eq} = 2^{1/6} \sigma$, $U_\text{harmonic}(r_{ij}) = \frac{1}{2} k (r_{ij} - r_\text{eq})^2$, where $k = 36 (2)^{2/3} \epsilon / \sigma^2$. In all simulations, atoms interact only with their nearest neighbors. This is a significant difference from other MD work. The main reason is that, since the harmonic potential does not tend to zero as $r_{ij} \to \infty$, it is ill suited to describe the forces of more distant neighbors. We limit the interactions within both potentials to include nearest neighbors only so that $U_\text{harmonic}$ approximates $U_\text{LJ}$ in a straightforward manner.

We have calculated the interfacial thermal conductance in systems with four different configurations of these forces: (a) all LJ, (b) all LJ except with harmonic interactions across the interface, (c) all harmonic except with LJ interface, and (d) all harmonic. Examples of steady-state temperature profiles from all four cases under otherwise identical simulation conditions are shown in Fig.~\ref{fig:Tz}. Each data point represents the average temperature in each bin as described in Appendix~\ref{sec:appendix_ar}, and the shaded region indicates the 95\% prediction interval for the bin temperatures. 
We note that the temperature profiles in cases (c) and (d) have effectively zero slope, corresponding to the diverging conductivity expected in a material with no phonon--phonon scattering.

\begin{figure}[t]
  \begin{center}
    \includegraphics{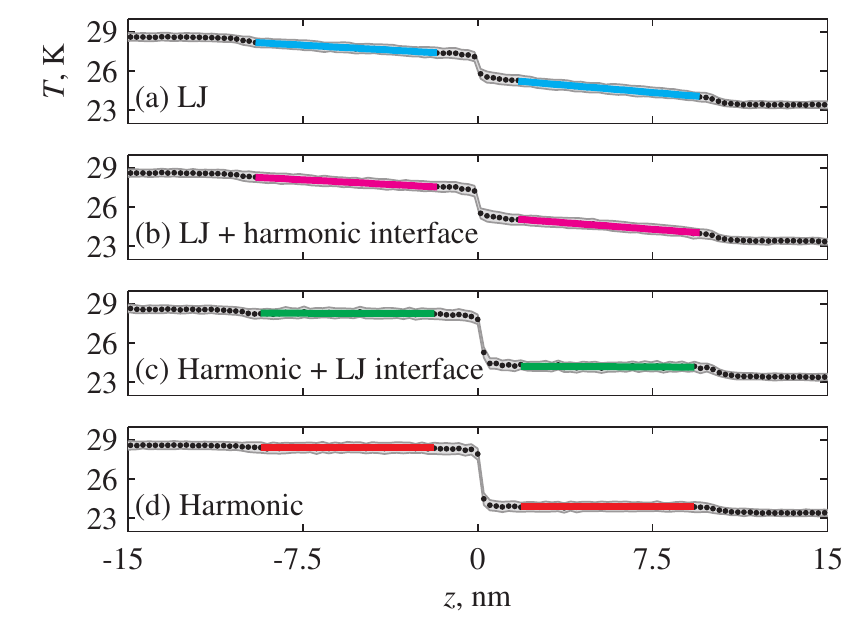}
    \caption{Steady-state temperature profiles in four identical systems except for the anharmonicity of the interatomic forces. The nominal temperature of these simulations is $T = 26$~K.}\label{fig:Tz}
  \end{center}
\end{figure}

\begin{figure}[t]
  \begin{center}
    \includegraphics{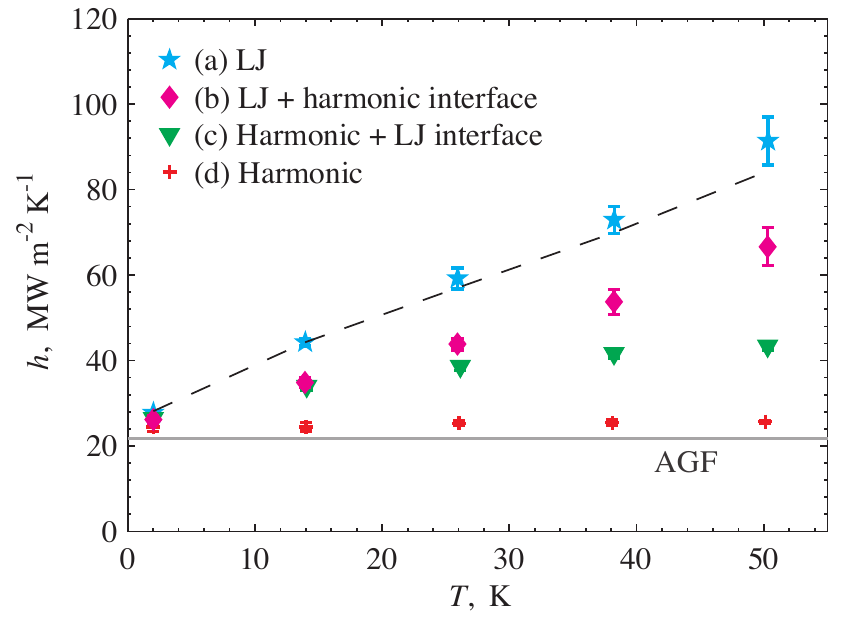}
    \caption{Thermal boundary conductance as a function of temperature in the same systems as in Fig.~\ref{fig:Tz}. The conductance calculated using atomistic Green's functions (solid gray) is expected to be similar to the conductance in the harmonic system. The dashed gray line is the sum of the harmonic conductance of case (d) with the ``excess'' conductances from cases (b) and (c).}\label{fig:h}
  \end{center}
\end{figure}

In order to calculate the interfacial thermal conductance from each simulation, the temperatures in the bulk leads are fitted to a linear profile and extrapolated to the interface, which allows the definition of the temperature drop $\Delta T$~\cite{English2012}. The conductance is then
\begin{equation}
  \label{eqn:h}
  h = \frac{\dot{Q}}{A \; \Delta T},
\end{equation}
where $A$ is the cross-sectional area and $\dot{Q}$ is the steady heat current added to the heat source and removed from the heat sink. Ten such simulations were performed in each system at each temperature with randomized initial velocities to provide independent trials. The mean conductance values from those trials are plotted in Fig.~\ref{fig:h} with error bars indicating 95\% confidence intervals. 

The conductances in all four systems converge at low temperatures, since displacements are small and the LJ potential is well approximated by the harmonic potential. As temperature increases, the conductance increases in case (a), as has been observed in MD simulations in previous work~\cite{Stevens2007,Landry2009a,Saaskilahti2014a}. The conductance also increases with temperature in cases (b) and (c), although at smaller rates. In contrast, the conductance in the harmonic case (d) is constant with temperature. These results are consistent with the hypothesis that, in the classical limit, increasing conductance with temperature is caused by inelastic processes, which are enabled by anharmonicity. The average and standard deviation of these values is $24.9 \pm 0.8$~MW~m$^{-2}$~K$^{-1}$. Empirically, we note that adding the ``excess'' conductance from cases (b) and (c) to the harmonic case (d) at each temperature produces conductance values (gray dashed line) very similar to those obtained in the all-LJ case (a). This lends support to the notion that the anharmonic contributions from the interface and from the bulk regions are simply additive.

For comparison, we have also calculated the conductance of the harmonic system using atomistic Green's functions (AGF)~\cite{Mingo2003a} as 21.69~MW~m$^{-2}$~K$^{-1}$, converged within 0.01~MW~m$^{-2}$~K$^{-1}$ with respect to wavevector and frequency sampling. Details of this calculation are given in Appendix~\ref{sec:appendix_negf}. There is a $\sim$13\% difference between the values obtained by AGF and by harmonic MD. We attribute the discrepancy to a difference in definitions of conductance between the two methods: AGF provides the total conductance between the two reservoirs, while NEMD provides the conductance at the interface between the leads. In other words, they differ in whether they include the additional resistance at the contacts with the external baths. Additional evidence is given in Appendix~\ref{sec:appendix_negf} to show that the contact resistance is responsible for the discrepancy. For our present purposes, the AGF calculation corroborates the magnitude of the conductance values from NEMD in the harmonic system, and provides additional evidence that our results are free of serious size and edge effects as cautioned in other work~\cite{Jones2013,Liang2014a}.

The key observation from Fig.~\ref{fig:h} is that the system consisting of LJ solids joined by harmonic interfacial forces [case (b)] exhibits a consistently higher conductance than the system of harmonic solids joined by LJ forces [case (c)]. Moreover, the discrepancy grows with temperature. We therefore conclude that, in this system, inelastic phonon processes in the bulk materials make a larger contribution to the conductance than inelastic processes at the interface.

\section{Role of Bulk Inelastic Scattering}
\label{sec:wavelet}

In this section, we present calculations of the energy distributions among the normal modes in the same NEMD simulations described in the previous section. By comparing the energy distributions, we elucidate the phonon phenomena that are responsible for the differences in conductance observed in Section~\ref{sec:nemd}. We use the wavelet transform, which has been applied previously to analyze the distribution of energy in MD simulations in spatial and spectral domains simultaneously~\cite{Baker2012}. To collect the signal to be transformed, we sampled atomic velocities every 40~ns during the same period in which the temperature profiles were collected. We chose to sample normal modes with wavevector $\mathbf{q}$ parallel to the $\langle 001 \rangle$ direction; therefore, we obtained the average velocity $\bar{\mathbf{v}}$ of atoms in each monolayer (\ie, each (002) plane) to form a one-dimensional signal $w_\alpha(z) = \sqrt{m(z)/2} \, \bar{v}_\alpha(z)$ corresponding to each Cartesian component $\alpha$. The wavelet transform of that signal, $\tilde{w}(z', q')$, is then used to calculate a kinetic energy density $E^\text{K}(z, q)$ as a function of both space and wavenumber. For ease of interpretation, we convert this to an equivalent temperature, $T_\text{equiv}(z, q)$; \ie, the temperature of a classical system at thermal equilibrium with an equal energy density. Details of this calculation are given in Appendix~\ref{sec:appendix_wavelet}. In principle, the same procedure can also be used to obtain the spectra of modes in directions other than (001) by sampling the corresponding planar velocities. However, the geometry of the system introduces complications in the interpretation of spectra in off-axis directions, so we present spectra along (001) only. For brevity, we also present only the spectra corresponding to longitudinal polarization; the spectra corresponding to transverse modes are very similar and lead to the same conclusions.

\begin{figure}[t]
  \begin{center}
    \includegraphics{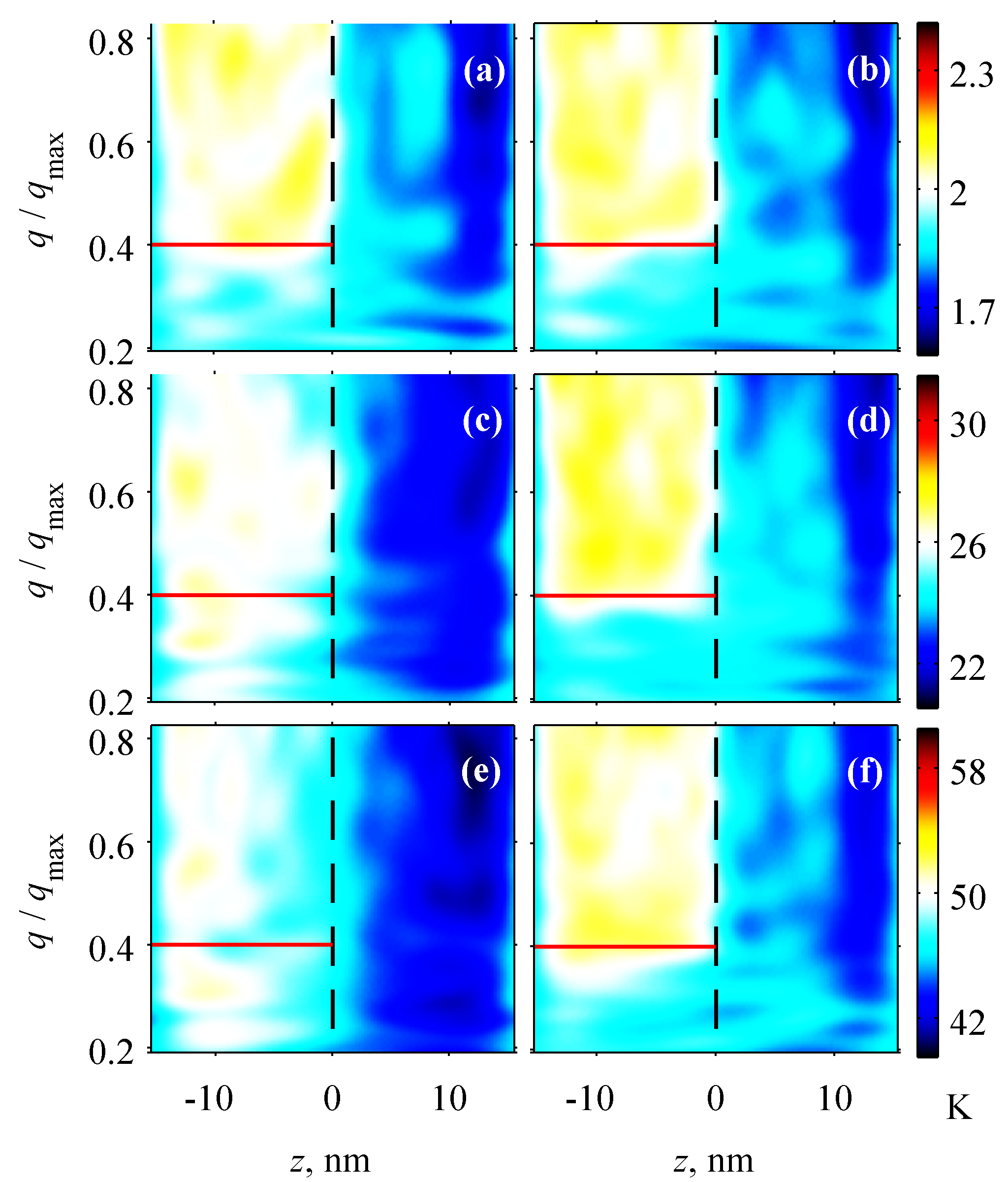}
    \caption{Distributions of kinetic energy in longitudinal $\langle 001 \rangle$ modes during the NEMD simulations of Section~\ref{sec:nemd} obtained by using the wavelet transform. The interface lies at $z = 0$ (dashed black), and modes with $q/q_\text{max} \approx 0.4$ in Ar (solid red) have the same frequency as the highest-frequency modes in heavy Ar. The nominal temperature increases from top to bottom: (a, b) 2~K, (c, d) 26~K, and (e, f) 50~K. The left panels (a, c, e) are calculated from systems with all LJ forces and the right panels (b, d, f) from systems with all harmonic forces.}
    \label{fig:wavelet1}
  \end{center}
\end{figure}

The resulting kinetic energy densities from six sets of NEMD simulations are plotted in Fig.~\ref{fig:wavelet1}. To reduce noise, the energy density shown in each panel is obtained from averaging those of ten identical simulations. Each paired row of panels is taken at the same temperature, increasing from top to bottom: (a, b)~2~K, (c, d)~26~K, and (e, f)~50~K. In each pair, the left panel is from an all-LJ system, and the right panel is from an all-harmonic system. As expected, the densities look similar at low temperature [panels (a) and (b)]. As temperature increases, the distributions diverge.

\begin{figure}[t]
  \begin{center}
    \includegraphics{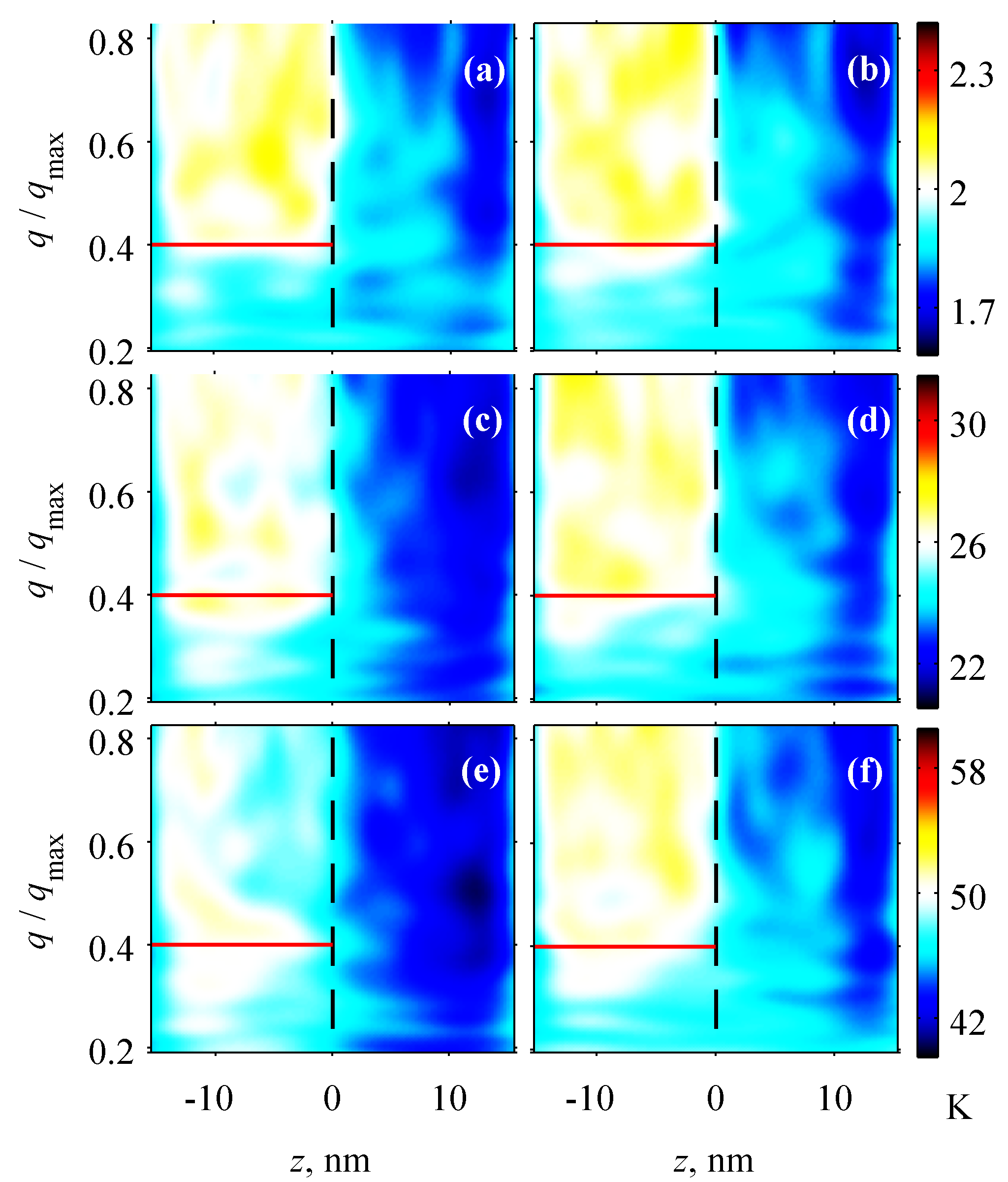}
    \caption{The same as Fig.~\ref{fig:wavelet1}, but the left panels (a, c, e) are calculated from LJ systems with harmonic interfacial forces and the right panels (b, d, f) are from harmonic systems with LJ interfacial forces.}
    \label{fig:wavelet2}
  \end{center}
\end{figure}

Interestingly, the energy distribution is in significant nonequilibrium on the Ar side in the LJ system at low temperature [panel (a)] and in the harmonic system at all temperatures [panels (b, d, f)]. Namely, there is excess energy in the modes with wavenumber above $q \approx 0.4 q_\text{max}$, while there is a deficit of energy below that threshold. The threshold coincides with the wavenumber of the Ar mode that has the same frequency as the cutoff frequency of heavy Ar, $q / q_\text{max} = 2 \pi^{-1} \sin^{-1}(m_\text{Ar}/m_\text{h-Ar})$. Therefore, we attribute the nonequilibrium to the fact that, in the harmonic system and in the low-temperature LJ system, phonons can only transmit elastically at the interface. High-frequency phonons originating in the Ar are therefore completely reflected at the interface, since there are no available modes of the same frequency in the heavy Ar.

As temperature increases in the LJ system, the atomic displacements increase, and the anharmonic forces enable the exchange of energy among modes of different frequency---\ie, the rates of inelastic processes increase. This leads to thermalization of vibrational energy in the Ar in the sequence from panel (a) to (c) to (e): the energy that is confined above $q/q_\text{max} \approx 0.4$ steadily relaxes into modes below the threshold. In light of the results of Section~\ref{sec:nemd}, this thermalization correlates with a drastic increase in thermal conductance of the interface. We therefore infer that conductance increases due to an increasing rate of thermalization of excess energy in high-frequency, non-transmitting modes to low-frequency modes with a high transmission.

The kinetic energy spectra of the remaining two types of systems are shown in Fig.~\ref{fig:wavelet2}: the left panels are from LJ systems with harmonic interfacial forces, and the right panels are from harmonic systems with LJ interfacial forces. In other words, the systems differ from those of Fig.~\ref{fig:wavelet1} only in the forces at the interface. The energy distributions of corresponding panels look remarkably similar, which implies that the interfacial forces have only a minor effect on the thermalization of modes in the Ar. In particular, we note that the LJ forces at the interface between harmonic solids only promotes thermalization very weakly if at all. This is associated with a relatively small increase in conductance with temperature in case (c) of Fig.~\ref{fig:h}, which we attribute to the bona fide interfacial inelastic phonon processes investigated in detail by S\"{a}\"{a}skilahti~\etal~\cite{Saaskilahti2014a}.

\section{Conclusions}
\label{sec:conclusions}


%
We have used classical molecular dynamics simulations to investigate the contributions to interfacial thermal conductance from anharmonic effects at the interface and in the nearby bulk materials. First, we confirmed that anharmonicity of interatomic forces is responsible for the increase of conductance with temperature. The results support the physically appealing model that the total thermal conductance at an interface is the sum of a contribution from elastic phonon transmission (which is constant in the classical limit) and a contribution from inelastic phonon processes that increases with temperature. We found that the inelastic part of the conductance can be further decomposed into contributions from bulk inelastic and interfacial inelastic processes. Between the two, the contribution from bulk inelastic processes is larger than that from the interface itself, and this difference grows with temperature. We then used the wavelet transform to obtain kinetic energy spectra, which demonstrate that the enhancement from bulk anharmonicity is due to an increased rate of thermalization of energy trapped in non-transmitting modes.

The present conclusions apply strictly to the Ar/heavy-Ar interface, which has been used extensively as a model system for interfacial thermal conductance. There are some aspects that should be investigated further to extend the findings to other systems, such as metal/diamond interfaces that have been measured experimentally. Like most other MD work on this problem, the materials are much more closely vibrationally matched than, e.g., the Pb/diamond interface; Gordiz and Henry recently investigated the effects of increasing mismatch explicitly, and showed that the anharmonic contribution to conductance becomes particularly important at large mismatch in bonding strength~\cite{Gordiz2015}. Nevertheless, the present findings provide important general guidance for the development of interfacial thermal conductance models that can accurately incorporate inelastic processes. Namely, our results show that it is not sufficient for conductance models to account only for frequency conversion at the interface, as done \eg\ in the HHIM~\cite{Hopkins2009d} and the AIM~\cite{Hopkins2011a}. In addition, it is necessary to account for the effective increase of incident phonon flux due to rethermalization of energy in modes with low transmission. New models that incorporate these effects should be able to take the parameters of the Ar/heavy-Ar system as input and test their predictions directly against the conductance contributions from elastic, bulk inelastic, and interfacial inelastic processes provided in this work.

\begin{acknowledgments}
N.Q.L., R.R., and P.M.N. acknowledge the financial support of the Air Force Office of Scientific Research (FA9550-14-1-0395). C.A.P., J.Z., and A.W.G. acknowledge financial support from NSF-CAREER (QMHP 1028883) and NSF-IDR (CBET 1134311). Computational work was performed using resources of the Advanced Research Computing Services at the University of Virginia and the Extreme Science and Engineering Discovery Environment (XSEDE) (DMR130123) supported by National Science Foundation (ACI-1053575). The authors are also grateful for useful discussions with C.H. Baker, P.E. Hopkins, and A.J.H. McGaughey.
\end{acknowledgments}

\begin{appendix}

\section{Ar/heavy Ar simulation details}
\label{sec:appendix_ar}

All molecular dynamics simulations were performed with the LAMMPS code package~\cite{Plimpton1995}. In choosing the model system, we sought the simplest system in which one can observe the effect of anharmonicity on thermal conductance and on phonon transport. A system meeting these criteria, similar to systems used in past MD studies of interfacial conductance~\cite{Stevens2007,English2012,Saaskilahti2014a}, is a coherent [001] interface between solid Ar (40~amu) and solid ``heavy Ar'' (120~amu). In this work, we use the LJ parameters $\epsilon = 0.01617$~eV and $\sigma = 3.347$~\AA, which correspond to the harmonic parameters $k = 0.8249$~eV\,\AA$^{-2}$ and $r_\text{eq} = 3.757$~\AA\@.

\begin{figure}[t]
  \begin{center}
    \includegraphics[width=3.4in]{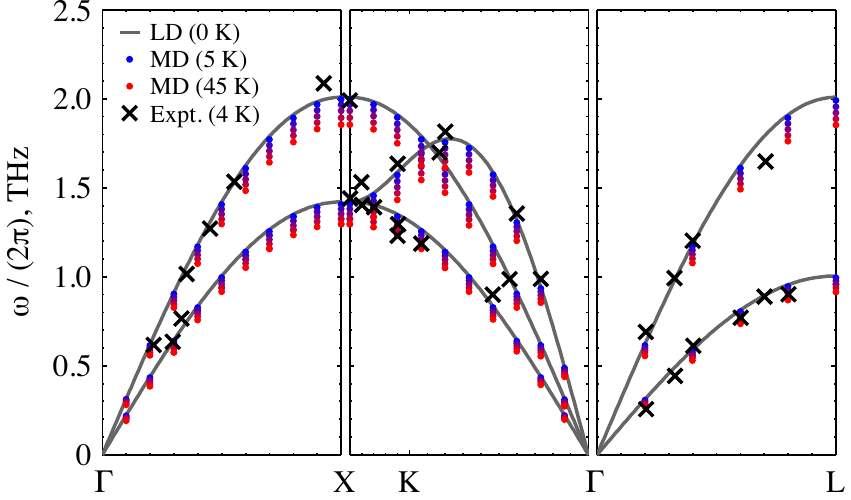}
    \caption{Dispersion of normal modes in simulated argon from lattice dynamics (LD) and from normal mode decomposition from molecular dynamics simulations (MD) compared with experimental measurements from Ref.~\onlinecite{Batchelder1970}.}\label{fig:dispersion}
  \end{center}
\end{figure}

These interatomic potentials produce a cubic lattice parameter of 5.313~\AA\ at 0~K, compared with 5.311~\AA\ extrapolated for Ar from experimental data~\cite{Peterson1966}. The potentials produce phonon dispersions in good agreement with neutron scattering measurements in solid Ar~\cite{Batchelder1970}, as shown in Fig.~\ref{fig:dispersion}. The highest-frequency mode has a vibrational period of 500~fs, based on which we select a timestep of 2~fs. To account for thermal expansion in the systems with LJ forces, simulations were performed to determine the zero-pressure lattice constant as a function of temperature. The simulations produced values of $a(T)$ that were fitted to a third-order polynomial function
\begin{equation}
  \label{eqn:aT}
  a(T) = a_0 + a_1 T + a_2 T^2 + a_3 T^3.
\end{equation}
The fitted coefficients are provided in Table~\ref{tab:aT}.

\begin{table}[t]
  \caption{Coefficients for Temperature-Dependent Lattice Parameter of LJ Argon [Eq.~\eqref{eqn:aT}]}
  \begin{center}
    \begin{tabular}{c r l}
      \hline
      Parameter & \multicolumn{2}{c}{Fitted Value} \\
      \hline
      $a_0$ & 5.313 & \AA\ \\
      $a_1$ & 1.813 $\times 10^{-3}$ & \AA\,K$^{-1}$ \\
      $a_2$ & 4.792 $\times 10^{-6}$ & \AA\,K$^{-2}$ \\
      $a_3$ & 1.394 $\times 10^{-8}$ & \AA\,K$^{-3}$ \\
      \hline
    \end{tabular}
    \label{tab:aT}
  \end{center}
\end{table}



The NEMD simulation domain has dimensions of $10 \times 10 \times 60$ conventional unit cells. The boundary conditions are periodic in the plane of the interface, approximating the interface between two slabs of infinite cross section. On each end, two (002) planes are held fixed as walls (400 atoms), and the temperature of the next twenty (002) planes (4000 atoms) is controlled using a Langevin thermostat with a time constant of 2.14~ps. S\"{a}\"{a}skilahti~\etal~\cite{Saaskilahti2014a} determined that this geometry was sufficiently large to avoid size effects in their system. Since (1) the forces in some of our systems are purely harmonic and (2) our LJ potential is limited to nearest-neighbor interactions, presumably reducing phonon--phonon scattering even in our anharmonic systems, we also performed additional simulations to check for size effects. Namely, we ran three series of simulations with increased cross-section ($15 \times 15$ cells), increased length (90 cells), and decreased thermostat time constants (1.07 and 0.54~ps) with no statistically significant change in conductance. The conductance from our AGF calculation (Section~\ref{sec:nemd} and Appendix~\ref{sec:appendix_negf}) also provides evidence that any size effects are not severe.


Each simulation began with the atoms in their equilibrium positions and with kinetic energy equivalent to twice the nominal temperature. For simplicity, the initial atomic velocities were set to the corresponding uniform magnitude of $|\mathbf{v}| = (2 d k_\text{B} T_\text{nominal} / m)^{1/2}$ with random orientation. The simulation then ran for 20~ps in order to reach thermal equilibrium. The thermostats were then applied at target temperatures of $(1 \pm 1/10) T_\text{nominal}$ for 4~ns, at which point we confirmed that the temperature distributions had reached steady state. To determine the temperature distribution, we divided the atoms into 120 bins along the transport direction, each bin containing one monolayer (200 atoms). The temperature was sampled in each bin in intervals of 1~ps. Running averages were stored in memory and written to disk every 40~ps, and those averages were collected for 8~ns, which provided 200 samples of the temperature in each bin.

\begin{figure}[t]
  \begin{center}
    \includegraphics{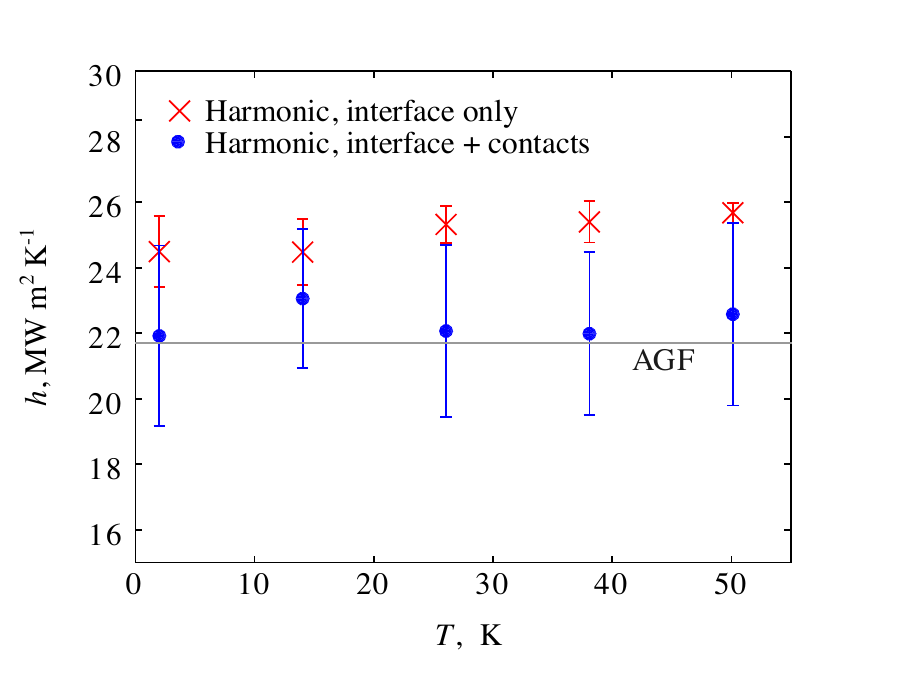}
    \caption{Conductance in the harmonic system calculated from NEMD using the lead temperatures extrapolated to the interface (red crosses) and the bath temperatures (blue circles), compared with the prediction from AGF.}\label{fig:contacts}
  \end{center}
\end{figure}

Each simulation thus provided a one-dimensional temperature distribution $T(z)$. We used an established procedure for extracting the thermal conductance at the interface: we fit a linear model to the temperature profiles in the two ``bulk-like'' regions and extrapolated them to the interface. We calculated $\Delta T$ as the difference between the extrapolated values, from which we calculated the conductance using Eq.~\eqref{eqn:h}. This conductance is physically different from the conductance calculated by atomistic Green's functions (AGF), since it excludes the small contact resistances between the two leads and the temperature baths. In Section~\ref{sec:nemd} we comment that this is the most likely explanation for the $\sim$13\% discrepancy between the NEMD conductance in the harmonic system and the AGF conductance. This is supported by re-calculating the NEMD conductance using the bath temperatures to calculate $\Delta T$ rather than the extrapolated temperatures at the interface. Those data are shown in Fig.~\ref{fig:contacts} (blue circles) in comparison with the conductance calculated in the usual manner (red crosses) and the AGF value (gray line). The inclusion of the contact resistances appears to be a plausible explanation for the discrepancy between the methods.

\section{Atomistic Green's Functions}
\label{sec:appendix_negf}

According to the formalism of atomistic Green's functions (AGF) in the harmonic limit, the thermal conductance across a device in between two contacts, each at thermal equilibrium, is given by~\cite{Carlos_Mingo2003,Carlos_Datta2005}
\begin{equation}
    h=\frac{1}{2 \pi A}\int\limits_0^\infty \hbar\omega \frac{\partial N}{\partial T} \text{Tr}\left\{\Gamma_l G \Gamma_r G^\dagger\right\} d \omega,
    \label{equG}
\end{equation}
where $A$ is the cross-sectional area, $\hbar \omega$ is the phonon energy, $N$ the Bose--Einstein distribution, and $T$ is the temperature. $G$ is the retarded Green's function for the dynamical equation of the device, which describes the response of the device's degrees of freedom upon an impulse excitation. $\Gamma_l$ ($\Gamma_r$) is the anti-Hermitian part of the left (right) contact self-energy. This quantity is related to the rate at which phonons leak from the device into the left (right) contact~\cite{Carlos_Datta2005}. Detailed explanations of the method and its numerical implementation are available in the literature~\cite{Carlos_Mingo2003,Carlos_Datta2005,Carlos_Zhang2007,Carlos_Wang2008,Carlos_Hopkins2009}; here we discuss details relevant to the present systems.

To compare the conductances calculated from AGF and classical MD simulations, we take the classical limit ($\hbar\omega \ll k_\text{B} T$) of Eq.~\ref{equG}. In that limit, the factor $\hbar \omega (\partial N / \partial T)$ reduces to the Boltzmann constant, $k_\text{B}$, and the thermal conductance becomes
\begin{equation}
    h = \frac{k_\text{B}}{2\pi A}\int\limits_0^\infty \text{Tr}\left\{\Gamma_l G \Gamma_r G^\dagger\right\} d \omega.
    \label{equGclassical}
\end{equation}

We used AGF to calculate the conductance at the Ar/heavy-Ar interface in the harmonic limit. The interatomic force constants were calculated from the Taylor expansion of the total energy, and we verified that they produce the same spectrum of normal modes. To calculate $\text{Tr}\left\{\Gamma_l G \Gamma_r G^\dagger\right\}$, we use the transverse symmetry of the system to decompose the problem into a sum of independent systems in the transverse $k$-space~\cite{Carlos_Zhang2006}. The transverse Brillouin zone was sampled with a grid of $200\times200$ equally spaced $k$-points.

\section{The wavelet transform}
\label{sec:appendix_wavelet}

The wavelet transform $\tilde{w}(q,z)$ of a signal $w(z)$ is an integral transform,

\begin{equation}
  \label{eqn:n_wavelets_W}
  \tilde{w}(z',q') = \mathcal{W} \{ w(z) \} = \int_{-\infty}^\infty w(z) \, \psi_{z',q'}^* (z) \, dz,
\end{equation}
where the kernel functions $\psi_{z',q'}$ are wavelets. We use the convention of Baker~\etal~\cite{Baker2012} in which each ``daughter wavelet,'' corresponding to a specific location $z'$ and wavenumber $q'$, is defined as

\begin{align}
  \label{eqn:n_wavelets_psi}
  \nonumber \psi_{z',q'}(z) = \pi^{-1/4} \, {\left( \frac{q'}{q_0} \right)}^{1/2} \, \exp \left[ i q' (z - z') \right] \, \times \\
  \exp \left[ - \frac{1}{2} {\left( \frac{q'}{q_0} \right)}^2 {(z - z')}^2 \right].
\end{align}
This is a scaled and translated version of a mother wavelet $\psi_{z',q_0}$ whose dominant wavenumber is $q_0$. The definition is normalized so that the energy density per length, per wavenumber is calculated as

\begin{equation}
  \label{eqn:n_wavelets_energy}
  E_\psi(z',q') = \frac{1}{C q_0} |\tilde{w}(z',q')|^2.
\end{equation}
We use the combination $w(z) = \sqrt{m(z)/2}\, v(z)$ as the signal to be transformed so that the wavelet energy density calculated by Eq.~\eqref{eqn:n_wavelets_energy} corresponds to the density of kinetic energy per length, per wavenumber. The constant $C$ accounts for the fact that, unlike the plane waves that form the basis functions for the Fourier transform, the wavelets are not orthogonal:

\begin{equation}
  \label{eqn:n_wavelets_C}
  C = \int_{-\infty}^{\infty} \frac{|\bar{\psi}_{z',q_0}(q)|^2}{|q|} dq,
\end{equation}
where $\bar{\psi}_{z',q_0'}(q)$ is the Fourier spectrum of the mother wavelet. We choose the wavenumber of the mother wavelet as $q_0 = 10/a$ and minimum and maximum wavenumbers corresponding to constants $\eta = 0.05$ and $\phi = 1$ as described in Ref.~\onlinecite{Baker2012}. These settings produce energy spectra with useful information in the range of wavenumbers between $q_\text{low} \approx 0.19 q_\text{max}$ and $q_\text{high} \approx 0.83 q_\text{max}$, which correspond to the limits on the vertical axes in Figs.~\ref{fig:wavelet1} and~\ref{fig:wavelet2}. To facilitate interpretation, the values plotted in those figures are not $E(z,q)$ itself, but rather the equivalent temperature
\begin{equation}
  \label{eqn:n_wavelets_T}
  T_\text{equiv}(z, q) = \frac{2 L_z (q_\text{high} - q_\text{low})}{k_\text{B}} E(z,q),
\end{equation}
where $L_z$ is the system length in the $z$ direction. That is, if a system were at thermal equilibrium with a uniform energy density of $E(z,q)$, then its temperature would be equal to $T_\text{equiv}(z, q)$.

\end{appendix}

\end{document}